\documentclass[aps,prb,preprint,groupedaddress,showkeys,showpacs]{revtex4}

\usepackage{bm}% bold math
\usepackage{graphicx}% Include figure files
\newcommand{\etal}{{\it et al} }

\newcommand{\RR}{\bm R}
\newcommand{\KK}{\bm K}
\newcommand{\qq}{\bm q}
\newcommand{\Aa}{\bm a}
\newcommand{\Zero}{\bm 0}
\newcommand{\Imath}{i}
\newcommand{\nWS}{n_{WS}}
\newcommand{\qEq}{\qq\!=\!\Zero}
\newcommand{\qNe}{\qq\!\ne\!\Zero}
\newcommand{\qKn}{\qq\!=\!\KK/n}

\begin{document}

%................. Titlepage ..............................

%\noindent {\verb/sv_15.tex/} \hfill 01/12/10 -- \today

\title
{
Renormalization method for infinite lattice sums revisited: 
lattice sums with Bloch phase factor
}

\author{\v{S}tefan Varga}\email{stefan.varga@savba.sk}

\affiliation{Institute of Inorganic Chemistry,
Slovak Academy of Sciences,
D\'ubravsk\'a cesta 9, SK-84536 Bratislava, Slovakia}

\date{\today}

\begin{abstract}

Infinite lattice summation scheme based on the idea 
of renormalization is generalized to enable evaluation 
of infinite lattice sums with Bloch phase factors which can occur 
when treating long-range interactions in infinite periodic systems. 
The scheme is fast, with easy to control accuracy and 
is not limited to any choice of special points in the Brillouin zone.
Illustrative calculation for a first few contributions 
for a simple cubic lattice is presented. 

\end{abstract}

\pacs{71.15.-m, 31.15.-p}
\keywords
{
infinite lattice sums, 
Bloch phase factor, 
renormalization method, 
solid harmonics
}

\maketitle

%..........................................................

%\Newpage

\section{Introduction}
\label{Introduction}

In electronic structure calculations of infinite or large finite systems 
the Laplace expansion of Coulomb potential ($R>a$) 
\begin{eqnarray}
\frac{1}{\vert\RR-\Aa\vert} = \sum_{l=0}^{\infty} \sum_{m=-l}^{l} 
{\cal I}_{lm}(\RR) {\cal R}_{lm}^{*}(\Aa)
% L_{lm}(\RR) M_{lm}(\Aa)
\label{MultExp}
\end{eqnarray}
is an indispensable frequently used factorization tool 
when treating distant interactions. 
In (\ref{MultExp}), $\cal R$ and $\cal I$ are the 
scaled regular and irregular solid harmonics, 
respectively,\cite{Helgaker} 
\begin{eqnarray}
{\cal R}_{lm}(\Aa) &=& \frac{1}{(l+m)!} \  a^l \  
P_{lm}(cos\theta_a) e^{\Imath m \phi_a},
\label{defR}
\\
{\cal I}_{lm}(\RR) &=& {(l-m)!} \  \frac{1}{R^{l+1}} \  
P_{lm}(cos\theta_R) e^{\Imath m \phi_R}
\label{defI}
\end{eqnarray}
%In (\ref{defI},\ref{defR}) 
$P$ are the Legendre polynomials and 
$\RR=(R,\theta_R,\phi_R)$, $\Aa=(a,\theta_a,\phi_a)$ in spherical coordinates.
%The scaling factors encountered in literature may differ (refs Kudin, white,) 
Apart from scaling factors 
%and possible complex conjugation 
%they 
%Eqs.(\ref{defR}--\ref{defI}) 
$\cal R$ and $\cal I$ 
correspond to 
%multipole and local moments\cite{White1994,Kudin2004} 
%used/appearing in fast multipole methods.\cite{Greengard1985} 
 multipole moments and their Taylor counterparts.\cite{White1994} 

In a number of calculations 
(e.g., post-Hartree-Fock 
%\cite{Sun1998,Kudin,Pisani} 
density fitting calculations in extended systems\cite{Pisani2007,Burow2017}) 
%reciprocal space formulation of HF and beyond in ext.systs, 
%density fitting in extended systems beyong Hartree approximation [Pis,Bur]) 
Bloch sums of (\ref{MultExp}) occur 
and the quality of long-range incorporation depends on how effectively 
one can cope with infinite lattice sums of the type 
\begin{eqnarray}
\sigma_{lm}(\qq) = \sum_{\RR>\RR_0}^{\infty} 
e^{\Imath \qq.\RR} {\cal I}_{lm}(\RR).
\label{siginf}
\end{eqnarray}
In (\ref{siginf}), 
$\RR_0$ indicates the short-range region of the infinite periodic lattice 
and $\qq$ is a wave vector. 

For one-dimensional periodicity (1D) infinite summations 
of the type (\ref{siginf}) do not present a problem. 
For $\qEq$ they lead to Riemann zeta functions, 
for  general $\qNe$ analytical formulas  
using Bernoulli numbers were derived.\cite{Monkhorst1988}  
Beyond 1D,   
for high $l$ one can profit from the fast decay of ${\cal I}(\RR)$  
which restricts the range of summation to a reasonable 
size. 
Anyhow, for the lowest orders of ${\cal I}$ 
brute force summations (if convergent at all) 
hardly return a sufficiently accurate answer in a reasonable time. 

For the $\qEq$ case, several efficient techniques were already published. 
The idea which dates back to Nijboer and De\,Wette\cite{Nijboer} is based on 
splitting the infinite sum into two parts using Ewald-like partitioning.  
Treating each part separately one gets a rapidly convergent direct space 
term and a fast converging term in reciprocal space. 
The idea was recast in a computationally more suitable form 
by Challacombe \etal\,.\cite{Chall1997}

An alternative way of treatment is based on a renormalization scheme. 
Looking at the multipole expansion at two different structural scales 
Bernman and Greengard\cite{Berman1994} succeeded in finding a relation 
between multipole expansion coefficients from which a recurrence formula 
for infinite lattice sums follows. 
The scheme was reinvestigated by Kudin and Scuseria\cite{Kudin2004} 
who formulated the idea of renormalization 
in terms of rescaling and translation operators. Their 
recurrence relation for infinite lattice sums is more natural to work with 
since each iterative step 
%has its direct geometric/structural meaning
%which can be monitored easily.
%represents directly 
can be interpreted as a contribution from certain part of the infinite lattice. 

The simple elegant idea behind the renormalization method, the fast convergence 
and implementational ease make the scheme challenging for trying to extend 
the technique to infinite sums of the type (\ref{siginf}) 
for nonzero $\qq$. 
When attempting to do so one has to find a way how 
--  in addition to profiting from the scaling properties of 
${\cal R}$,\,${\cal I}$  --  
to cope with the Bloch phase factor scaling.
To the author's knowledge the only attempt to generalize the scheme 
of Kudin and Scusseria\cite{Kudin2004} beyond $\qEq$ 
was published by Grundei and Burow 
(Appendix A of Ref.\,\onlinecite{Burow2017}) 
where the authors simply got rid of the phase scaling problem 
for the price of limiting their scheme to the restriction to  
$\qq$ being a fractional part of a reciprocal lattice vector, $\qKn$. 
Selecting $\qq$ in this special form  
for a suitable choice of the initial cluster always returns  
an integer multiple of $2\pi$ phase and 
the phase scaling problem does not occur. 
We show that an easy way how to solve the problem for completely general 
$\qq$ without any restriction to special type of wave vector 
exists for only modest additional costs.

In Sec.\,\ref{Method} we review the ideas of renormalization method 
in easy to follow geometric way first. Using this picture we derive 
the recurrence formula for fast evaluation of lattice sum (\ref{siginf}) 
for general nonzero wave vector 
which does not suffer from any restrictions imposed on $\qq$. 
In Sec.\,\ref{Discussion} numerical issues are discussed and an illustrative 
calculation for a simple cubic lattice is presented.

\section{Method}
\label{Method}

The basic idea behind the technique is the same as 
for the $\qEq$ case,\cite{Kudin2004}  
however, our way of derivation is free of involved operator manipulations,  
it offers a simple geometric insight 
and enables an easy generalization to the nonzero wave vector case. 

To understand the essence of the method let us describe the way how the 
infinite lattice is generated first.
We will consider a system with translational periodicity 
in three dimensions (3D). 
By a straightforward simplification, 
the scheme can be applied to periodicity in one or two dimensions as well.  

The idea is illustrated in Fig.\,1.
A cluster consisting of the unit cell centered at the origin 
and its first $\nWS$ neighbour shells 
($\nWS$ is the analogy of the well-separatedness 
criterion\cite{Greengard1985,White1994})  
%used/introduced in Ref.\,White94,\cite{Greengard1985}) 
we will refer to as the central cluster (CC).  
We will distinguish the CC (with a general integer $\nWS>0$) 
from a special case of CC with $\nWS=1$ 
(always of size $n_0=3\!\times\!3\!\times\!3$ cells) 
which we will call the nearest neighbours cluster (NNC).

Consider now a layer of cells next to CC consisting of all nearest neighbour 
replicas of CC (its edge being 3-times the edge of CC). 
We denote the number of cells in this layer by $N_0$ (Layer I in Fig.\,1). 
Let us label each cell by its lattice translation vector.  
When we stretch the lattice translation vector $\RR_0$ associated with a cell 
from Layer I by a factor of 3 we get a new lattice translation vector 
$\RR'_0\!=\!3 \RR_0$ which is now a center  of $ 3\!\times\!3\!\times\!3$ 
supercell (a periodic replica of NNC) in the next layer (Layer II in Fig.\,1). 
When we repeat the process with all $\RR_0$ from Layer I we end up with 
a completely filled Layer II.  
Evidently, there are $N_0$ supercells of $n_0$ cells each in Layer II 
and
the complete set of lattice translation vectors of all cells from Layer II 
consists of $N_1\!=\!N_0\!\times\!n_0$ vectors 
$\RR_1\!=\!\RR'_0\!+\!\Aa\!=\!3\RR_0\!+\!\Aa$, 
where $\Aa$ runs over all $n_0$ lattice translation vectors of the NNC. 

This process can now be repeated recursively until sufficiently large cluster 
is generated. From the way of construction 
it follows that the relation between sizes of two successive layers will 
always be $N_{n+1}\!=\!N_n\!\times\!n_0$ so that the size of a layer 
exhibits geometric growth.  
%grows as fast as a geometric progression 
Consequently, large enough cluster can be generated within 
a small number of recursive steps in this way.

Let us return to the evaluation of infinite sums (\ref{siginf}) now. 
Decomposing the infinite sum into contributions from all layers we have
\begin{eqnarray}
\sigma_{lm}(\qq) = \sigma_{lm}^{(0)}(\qq) + \sigma_{lm}^{(1)}(\qq) + \dots +  
\sigma_{lm}^{(n)}(\qq) + \dots
\label{sigall}
\end{eqnarray}
where
\begin{eqnarray}
\sigma_{lm}^{(n)}(\qq) = \sum_{\RR_n} e^{\Imath \qq.\RR_n} {\cal I}_{lm}(\RR_n)
\label{sign}
\end{eqnarray}
and the summation in (\ref{sign}) runs over all the $N_n$ 
lattice translation vectors 
$\RR_n$ of the n-th layer.
Let us suppose we already know $\sigma_{lm}^{(n)}(\qq)$. 
Considering the way how layers were constructed the contribution 
from the next layer will be 
\begin{eqnarray}
\sigma_{lm}^{(n+1)}(\qq)
\!&=&\!\sum_{\RR_{n+1}} e^{\Imath \qq.\RR_{n+1}} \, {\cal I}_{lm}(\RR_{n+1}) 
\nonumber
\\
&=&\!\sum_{\RR_n} \sum_{\Aa} e^{\Imath \qq.(3\RR_n+\Aa)} \, 
{\cal I}_{lm}(3\RR_n+\Aa) 
\nonumber
\\
&=&\!\sum_{\RR_n} e^{\Imath \qq.3\RR_n} \, 
\sum_{jk} {\cal I}_{l+j,m+k}(3\RR_n) \,
\sum_{\Aa}  e^{\Imath \qq.\Aa} \, {\cal R}_{jk}^{*}(-\Aa)
\nonumber
\\
&=&\!\sum_{jk} \xi_{l+j} \, 
 \sum_{\RR_n} e^{\Imath 3\qq.\RR_n} \, 
{\cal I}_{l+j,m+k}(\RR_n) M_{jk}^{*}(\qq).
\label{signext}
\end{eqnarray}
%
%In (\ref{signext}) 
Use was made of the 
%factorisation property of $L$ (see Eq.\,(17) in Ref.\,\onlinecite{White1994}).
addition theorem for irregular solid harmonics\cite{Caola1987} 
which in normalization (\ref{defR}--\ref{defI}) 
reads
\begin{eqnarray}
{\cal I}_{lm}(\RR-\Aa) = \sum_{j=0}^{\infty} \, \sum_{k=-j}^{j} \, 
{\cal I}_{l+j,m+k}(\RR) \, {\cal R}_{j,k}^{*}(\Aa). 
\label{Iaddtheorem}
\end{eqnarray}
%
%The factor $\xi_l=3^{-(l+1)}$ 
The factor $\xi_l=1/3^{l+1}$ is a result of scaling property of ${\cal I}$.  
Comparison with (\ref{sign}) yields the recurrence formula
\begin{eqnarray}
\sigma_{lm}^{(n+1)}(\qq) = \sum_{j=0}^{\infty} \, \xi_{l+j} \sum_{k=-j}^{j} \, 
\sigma_{l+j,m+k}^{(n)}(3\qq) \ M_{jk}^{*}(\qq),
\label{sigrecurr}
\end{eqnarray}
or, in more compact form using the symbolics of Ref.\,\onlinecite{Kudin2004},
\begin{eqnarray}
\sigma^{(n+1)}(\qq) = {\cal U}_L [\sigma^{(n)}(3\qq)] \otimes M^*(\qq).
\label{Symb_sigrecurr}
\end{eqnarray}
In (\ref{sigrecurr}) $M(\qq)$ was introduced
\begin{eqnarray}
M_{jk}(\qq) = 
\sum_{\Aa} \, e^{\Imath \qq.\Aa} \, {\cal R}_{jk}(\Aa)
\label{MNNC}
\end{eqnarray}
where the summation runs over all $n_0$ lattice translation vectors of the NNC.
The recurrence relation (\ref{Symb_sigrecurr}) is the key formula of our scheme.
As expected, for $\qEq$ it simplifies to 
a formula equivalent to Eq.\,(24) of Ref.\,\onlinecite{Kudin2004}, 
the formal difference is that we evaluate the layer contributions 
while in Ref.\,\onlinecite{Kudin2004} the partial sums are treated directly.

\section{Calculational details and Discussion} 
\label{Discussion}

Similar to the $\qEq$ scheme\cite{Kudin2004} 
once we know $\sigma^{(0)}$ and $M$ 
all the contributions to full $\sigma$ can be evaluated recursively 
using (\ref{Symb_sigrecurr}). Notice, however, that for knowing e.g.
$\sigma^{(1)}(\qq)$ we need to know $\sigma^{(0)}(3\qq)$, etc. 
Consequently, to get $\sigma^{(n)}(\qq)$ we need to start the recurrence from 
$\sigma^{(0)}(3^n\qq)$. 
At the same time, 
$M(3^m\qq)$ will also be required, $m=0,1,\dots,(n-1)$. 
This is the additional expense we have to pay in our scheme.  
Notice that there is no danger of numerical overflow for any $3^n\qq$ \,
since -- owing to periodicity of $\sigma(\qq)$ in reciprocal space -- each 
wave vector can be always kept within the first Brillouin zone (BZ) 
by a suitable reciprocal lattice translation.

Compared with the $\qEq$ case general nonzero $\qq$ calculation 
converges considerably faster. 
Typically, 8-9 iterations are sufficient for 16 digit accuracy 
for a general $\qq$-vector from inside BZ for $\nWS\!=\!1$ and $l\!\geq\!3$.
For special points at BZ edge number of iterations 
varies between 7 for $\qq\!=\!(1/2,1/2,1/2)$ and 11 for $\qq\!=\!(1/2,0,0)$ 
(in reciprocal lattice vector units)
while for $\qEq$ (and for any other reciprocal lattice vector) 
16 iterations are needed for 16 digit convergence. 
%For very small $\qq$ number of iterations increases 
For small $\qq$ in symmetry positions number of iterations can be 
similar to the $\qEq$ case.

Other numerical issues behave in a way similar to that in 
Ref.\,\onlinecite{Kudin2004}. 
The infinite summation over angular momentum in (\ref{sigrecurr}) 
has to be truncated to some finite value $l_{max}$ in practice. 
We found $l_{max}\!=\!40$ sufficient for all $\sigma_{lm}$ to be saturated. 
It is natural to require to have also large enough $l_{max}$ 
for which the results are already not sensitive to the choice of $\nWS$, 
%large enough for $\nWS$ to have no effect on results, 
i.e., for which the computationally least demanding choice 
$\nWS\!=\!1$ is sufficient. 
To be able to compare the effect of the choice of $\nWS$ 
we unified $\RR_0$ in (\ref{siginf}) 
for calculations with different $\nWS$ first. 
We set $\RR_0\!=\!\Zero$ and 
the missing finite part beyond $\RR\!=\!\Zero$ was 
added to each $\sigma_{lm}$ 
to make the comparison for different $\nWS$ possible. 
In all cases, the differences between $\nWS\!=\,1$ and $\nWS\!=\!2$ 
were below 15th decimal place  
for $l_{max}=40$. 

Working in double precision arithmetics, 
the symmetry-expected zero imaginary (or, real) parts of infinite sums 
we found all to be within $10^{-15}$ accuracy of corresponding 
$\vert\sigma\vert$. 
The zero-to-be sums were also within 15 digit accuracy compared with 
nonzero terms of the same $l$ (or, close $l$ if $\sigma_l=0$ all). 
Of course, for large $l$ (where $\vert\sigma_l\vert\gg\!1$) 
we can get spurious numbers orders of magnitude above $10^{-15}$ 
instead of true zeros.   
If wishing to get rid of these artifacts Legendre polynomials have to be 
evaluated in quadruple precision arithmetics.

%\include{Tab1}
%Table.
%\begin{table*}
\begin{table}
\caption
{
First few lattice sums (\ref{siginf}) 
with $\RR_0\!=\!\Zero$ for $\qq\!=\!(0.1,0.1,0.1)$  
[reciprocal lattice vector units] for a simple cubic lattice, $\nWS\!=\!1$. 
}
\begin{ruledtabular}
\begin{tabular}{rrrr}
 $l$  &  $m$  &  
    \multicolumn{1}{c}{$Re\{\sigma_{lm}(\qq)\}$}  &  
    \multicolumn{1}{c}{$Im\{\sigma_{lm}(\qq)\}$} \\
\colrule
    3  &  0  &         0.000000000000000  &         7.461180731804426 \\
    3  &  1  &        -3.730590365902213  &         3.730590365902213 \\
    3  &  2  &         4.575050777090631  &         0.000000000000000 \\
    3  &  3  &       -18.652951829511067  &       -18.652951829511067 \\
    4  &  0  &        60.616977645071896  &         0.000000000000000 \\
    4  &  1  &         0.907115047217165  &         0.907115047217165 \\
    4  &  2  &         0.000000000000000  &         3.628460188868663 \\
    4  &  3  &         6.349805330520163  &        -6.349805330520163 \\
    4  &  4  &       303.084888225359527  &         0.000000000000000 \\
    5  &  0  &         0.000000000000000  &       120.923466268988799 \\
    5  &  1  &        51.788023546202568  &       -51.788023546202568 \\
    5  &  2  &         0.000000000000001  &         0.000000000000000 \\
    5  &  3  &        86.482861899369858  &        86.482861899369858 \\
    5  &  4  &         0.000000000000000  &        51.533789562654259 \\
    5  &  5  &       984.480915344945550  &      -984.480915344945550 \\
\end{tabular}
\end{ruledtabular}
\label{qxyz}
\end{table}
%\end{table*}

In Table\,\ref{qxyz}
lattice sums starting from $l\!=\!3$  
(the lowest order free of possible conditional convergence for 3D system) 
up to $l\!=\!5$  
are presented for 
%lattice sums up to $l\!=\!6$ are presented for 
$\qq\!=\!(0.1,0.1,0.1)$ 
%\ref{qedge},
%the nonzero lattice sums up to $l\!=\!12$ are presented for 
%$\qq\!=\!(0,0,1/2)$ and $\qq\!=\!(1/2,1/2,1/2)$ 
for a simple cubic lattice with unit lattice constant.
For all numbers in Table\,\ref{qxyz} 
Legendre polynomials were evaluated in quadruple precision arithmetics.

Compared to $\qEq$ case our algorithm is slowed down by 
the fact that for each  
$\sigma^{(n)}(\qq)$ sets of $\sigma^{(0)}(3^n\qq)$ and $M(3^{n-1}\qq)$ 
have to be evaluated for each $\qq$. 
The special $\qq$-points scheme\cite{Burow2017} 
does  not suffer from these extra expenses either, 
however, 
in their scheme the choice of central cluster is governed by the choice 
of $n$ in $\qKn$,   
%Not all fractional $\qq$-points are easily available in their scheme, 
which for some $\qq$ makes the evaluations in the central cluster 
%can turn out more 
quite demanding. 

In our scheme, each $\qq$-point of our need we get for 
the same price without any additional complication. 
Moreover, since the sums (\ref{siginf}) are sufficient to be pre-calculated 
at the initial stage once per calculation 
(and can be used for the same lattice with the same choice of $\qq$-points 
repeatedly) extreme speed of the summation scheme is not an issue that 
should bother us. The possibility to do the summation  
with any $\qq$ without any restriction in a reasonable time 
with a controllable accuracy is what is usually needed. 
%And that is what our scheme conveniently does.

In spite of the need to start the recurrence from a new $\sigma^{(0)}(3^m\qq)$ 
and have all $M(3^{m-1}\qq)$ available in each $m$-th recurrence  
the algorithm can be arranged so 
that the computational costs scale linearly with the number of 
recursive steps. At the same time, 
number of terms included in summation grows geometrically.

\section{Conclusions}
\label{Conclusions}

We generalized the renormalization idea based 
lattice summation method of Kudin and Scuseria\cite{Kudin2004} 
to enable evaluation of 
infinite lattice sums with Bloch factor. 
The scheme is general 
and is not limited to any special form of the wave vector. 
As a by-product, we offer a simple novel way 
of looking at the renormalization scheme.

For a general point from inside the Brillouin zone the number of recurrence 
steps is typically lower than for the $\qEq$ case.
Compared to the $\qEq$ scheme\cite{Kudin2004} the method has only modest 
extra computational expenses.

\begin{acknowledgments}

This work was supported by the Slovak grant agency VEGA
(Project No.\,2\,-\,0116\,-17).
Discussions with M\'arius K\'adek 
who brought the renormalization method to the author's attention 
are acknowledged.

\end{acknowledgments}

\bigskip

\bigskip

\begin{picture}(260,160)(-80,-80)

\linethickness{1.25pt}
%CC
\put(-20,-20){\line(1,0){40}}
\put(-20,-20){\line(0,1){40}}
\put( 20, 20){\line(-1,0){40}}
\put( 20, 20){\line(0,-1){40}}
%Layer I
\put(-60,-60){\line(1,0){120}}
\put(-60,-60){\line(0,1){120}}
\put( 60, 60){\line(-1,0){120}}
\put( 60, 60){\line(0,-1){120}}
%Layer II
\put(180,80){\line(0,-1){160}}
%R0
\put(44,12){\line(1,0){8}}
\put(44,12){\line(0,1){8}}
\put(52,20){\line(-1,0){8}}
\put(52,20){\line(0,-1){8}}
%R0'
\put(132,36){\line(1,0){24}}
\put(132,36){\line(0,1){24}}
\put(156,60){\line(-1,0){24}}
\put(156,60){\line(0,-1){24}}

\thicklines
%\put(0,0){\vector(3,1){48}}
\put(0,0){\vector(3,1){50}}
\put(0,0){\vector(3,1){144}}

%\thinlines
\linethickness{0.25pt}
%cells
\multiput(-76,-80)(8,0){32}{\line(0,1){160}}
\multiput(-80,-76)(0,8){20}{\line(1,0){260}}

%labels
\put(  0,-15){\makebox(20,20)[lb]{{\it{CC}}}}
\put( 15,-50){\makebox(40,20)[lb]{{\it{Layer I}}}}
\put(120,-70){\makebox(40,20)[lb]{{\it{Layer II}}}}
\put(100, 40){\makebox(20,15)[lb]{$\RR'_0$}}
\put( 28, 22){\makebox(20,15)[lb]{$\RR_0$}}

\end{picture}

\bigskip

Fig.\,1. 
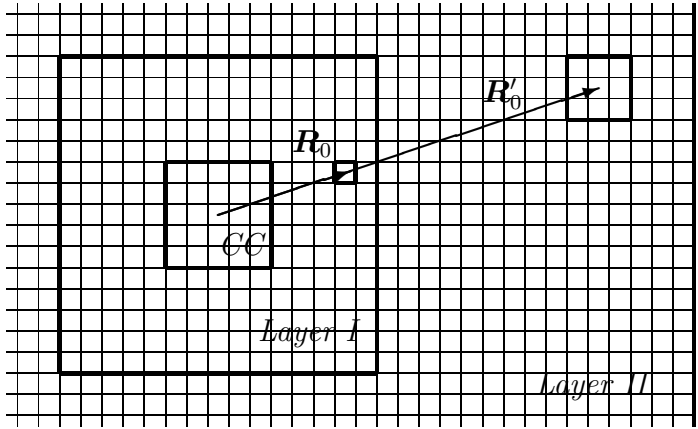
Two-dimensional illustration of lattice generation for $\nWS=2$: 
the central cluster, Layer I and a part of Layer II. 
$\RR'_0$ is the $3\times$ stretched $\RR_0$ lattice translation vector; 
the $\RR_0$-cell and the corresponding supercell around $\RR'_0$ 
are set off by bold framing.

\end{document}